\documentstyle[psfig]{lamuphys}
\makeatletter
\let\chapter\hid@chapter
\makeatother
\begin{document}
\pagenumbering{arabic}
\title{Bimodality of Freeman's Law}

\author{R. Brent\,Tully\inst{1}, and Marc A.W.\,Verheijen\inst{2}}

\institute{Institute for Astronomy, University of Hawaii\\
2680 Woodlawn Drive, Honolulu, Hawaii  96822, USA
\and
Kapteyn Astronomical Institute\\ Postbus 800, NL-9700 AV
Groningen, The Netherlands}

\maketitle

\begin{abstract}
A cluster sample of 62 galaxies complete to $M_B=-16.5^m$ has been
observed at $B,R,I,K^{\prime}$ bands with imaging detectors.  The
distribution of exponential disk central surface brightnesses is found
to be bimodal.  The bimodality is particularly significant at
$K^{\prime}$ because obscuration is not a problem and because the high
surface brightness galaxies are redder than the low surface brightness
galaxies so the bifurcation is greater.  The bimodality signal is
especially clear when galaxies with close companions are excluded from
consideration.  High and low surface brightness pairs with essentially
identical luminosities and maximum rotation characteristics are
compared.  It is suggested that the high surface brightness galaxies
have self-gravitating disks while the low surface brightness galaxies
are halo dominated at all radii.  Evidently the intermediate surface
brightness regime is unstable.  If a disk has sufficiently low angular
momentum and it shrinks enough that the disk potential begins to
dominate the halo potential locally, then the disk must secularly
evolve to the high surface brightness state characterized by a flat
rotation curve.
\end{abstract}
\section{Introduction}
There has been a debate about whether the disks of galaxies have a
quantized central surface brightness (\cite{fr}; \cite{vdk}) or,
instead, there is a broad distribution of disk central surface
brightnesses (\cite{mc:bo:sch}; \cite{dj}).  \cite{di} and others have
discussed the selection effects that could distort the observed
surface brightness distribution.  Surveys of the brightest galaxies
tend to be dominated by the highest surface brightness objects.
Special efforts are required to find low surface brightness galaxies
(\cite{vdb}; \cite{im:bo:ma}; \cite{da}).  It is difficult to have a
proper normalization of the relative occurance rates in different
surface brightness intervals.

A large and complete sample of galaxies are
considered in this study.  The galaxies are drawn from the Ursa Major
Cluster and the cluster assignment requires an appropriate redshift.
The cluster is dynamically 
young (long crossing time, no central concentration, no ellipticals,
many HI-rich spirals) and the galaxies may be representative of those
in low
density environments.  There are 62 galaxies in the complete sample
with $M_B<-16.5^m$.  The data are presented in \cite{tu:ve}.

CCD photometry at $B,R,I$ bands has been acquired for all of the
complete sample and $K^{\prime}$ HgCdTe images have been obtained for
60 of 62.  This $K^{\prime}$ data is a key to the discussion.
First, the $K^{\prime}$ material is not affected significantly by
obscuration so surface brightnesses are only modified by geometric
effects associated with viewing perspectives.  Second, higher
surface brightness galaxies tend to be redder than lower surface
brightness galaxies so the distinctions between the different classes
that will be discussed are clearer as one progresses to the red.

\section{Bimodality of Central Surface Brightnesses}
There is a more complete presentation by \cite{tu:ve:2}, where the
situation in all available passbands is discussed.  Only the best
evidence will be given here.  The bimodality is found in all the
observed bands
but is the least ambiguous at $K^{\prime}$.  In addition, it turns out
that the few intermediate surface brightness cases in our sample all
have close companions.  If the galaxies with substantial companions
closer than 80~kpc in projection are excluded, the distribution of
surface brightnesses of the remaining 36 of 60 galaxies with
$K^{\prime}$ photometry is that seen in Figure~1.

\begin{figure}
\centerline{
\psfig{figure=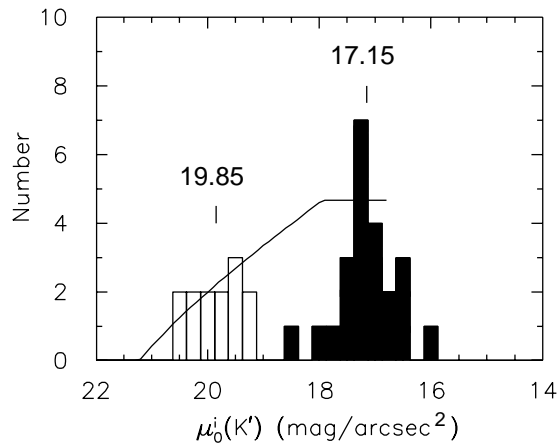,height=6cm}
}
\caption{Histogram of exponential disk central surface brightnesses at
$K^{\prime}$.  Inclination corrected assuming no obscuration.
Isolated galaxies only; galaxies with companions within 80 kpc
projected are excluded.  Filled histogram: HSB galaxies; open
histogram: LSB galaxies.  Solid curve: completeness expectation if
there are equal numbers of galaxies in all surface brightness bins
$>17$ mag/sq. arcsec.  Means of the HSB and LSB distributions are
indicated.}
\end{figure}

There is a clear separation into two peaks.  The filled part of the
histogram illustrates the $K^{\prime}$ equivalent of `Freeman's law'.
The galaxies that contribute belong to a {\it high surface brightness}
[HSB] family.  The open part of the histogram is made up of
constituents of a {\it low surface brightness} [LSB] family.  The
superimposed solid curve illustrates a completeness expectation
assuming there is uniform population of the surface
brightness--exponential scale length domain for scale lengths between
500~pc and 3~kpc and for all surface brightnesses faintward of
17~mag/sq.~arcsec.  It can be seen that the LSB peak has probably been
reduced by incompletion and may well have been truncated on the faint
side.  However, the critical point is that the visibility is quite
adequate for the
intermediate surface brightness regime between the HSB and LSB peaks
and the paucity of objects in that region is highly significant. 

\section{Luminosity--Line Width and Surface Brightness--Scale Length
Diagrams} 
Galaxies with essentially identical luminosities and rotation maxima
can have very different central surface brightnesses and exponential
scale lengths.  These properties are demonstrated in Figure~2.  The
left panel shows the tight relation between $K^{\prime}$ luminosities
and HI line profile widths.  The line profiles have been transformed
to approximate rotation maxima.  HSB galaxies are represented by
filled symbols and LSB galaxies, by open symbols.  Three pairs are
identified by numbers.  The triplet identified by the letter $a$ is 
two HSB galaxies, one with, and one without, a bulge and an LSB.  
The distinctions between bulge
and non-bulge systems are discussed by \cite{tu:ve:2}.

The right panel illustrates the relation between $K^{\prime}$ central
disk surface brightness and exponential scale lengths.  Surface
brightnesses have been corrected for inclination projection effects
assuming no obscuration.  Filled and open symbols have the same
meaning as in the other panel.  Crosses locate galaxies too faint to
be in the complete sample.   There are fewer data points in the left
panel because only objects with satisfactory types, HI information,
and inclinations could be plotted.

\begin{figure}
\centerline{
\psfig{figure=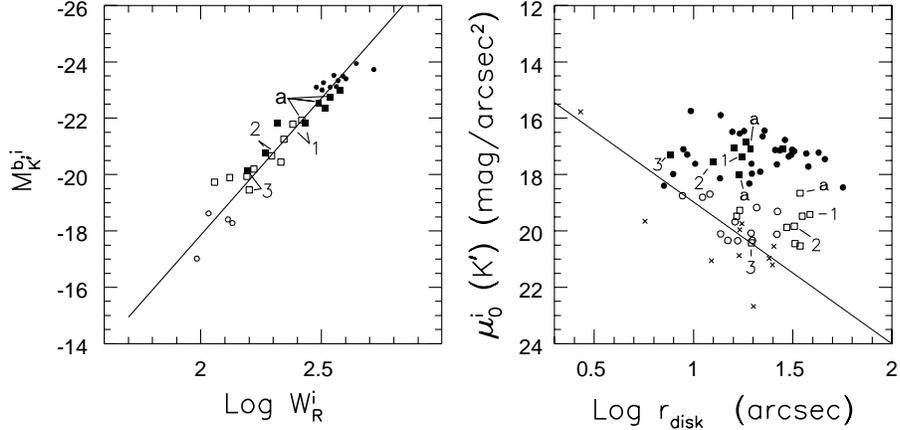,height=6cm}
}
\caption{{\it Left panel:}  Luminosities at $K^{\prime}$ vs. HI
profile line width; inclination corrected.  Filled symbols: HSB; open
symbols: LSB.  HSB--LSB pairs with similar luminosities and linewidths
are labeled 1,2,3.  A bulge--nonbulge HSB pair and an LSB of comparable
luminosity are labeled $a$.  The
solid line is the regression with uncertainties in line widths.
{\it Right panel:} $K^{\prime}$ disk central surface brightness
vs. exponential scale length.  Symbols as in left panel with the
addition that crosses are galaxies fainter than the completion limit.
The same pairs and triplet are labeled.  The solid line is the 
approximate completion limit.}
\end{figure}

It is immediately seen how the HSB and LSB systems that were paired in
the left panel have moved to separate domains in the right panel.
Galaxies with a given total amount of light can be bright and
compressed or dim and extended, yet generate the same rotation
speeds.  Galaxies apparently avoid an intermediate region in the
parameter space of the right panel.  The one-dimensional projection of
the right panel of Fig.~2 onto the surface brightness axis gives the
equivalent of Fig.~1 for the entire sample.  The few intermediate
surface brightness cases in Fig.~2 are filtered out of Fig.~1 because
they are in close proximity to another galaxy.

\section{Rotation Curve Disk--Halo Decompositions}
Velocity field information will be made available for a large fraction
of the Ursa Major sample (\cite{ve}) but that work is still in
progress.  An HSB--LSB luminosity--rotation maximum pair that 
is not part of the Ursa Major sample
has been discussed in the literature.
\cite{db:mc} have decomposed the rotation
curves of the HSB galaxy NGC~2403 and the LSB galaxy UGC~128.
Their results are consistent with the following discussion.

At present, we consider only the three galaxies identified by the 
labels $a$ in Fig.~2.  One of these, NGC~3917, is an example of the 
LSB family.  The other two are HSB examples: NGC~3949 has an 
exponential disk but no substantial bulge while UGC~6973 has a 
central concentration of light in addition to an exponential disk.
Figure~3 provides a graphic summary of much of our information about 
these systems.  There are $B$ band images, I band surface brightness 
profiles, HI position-velocity maps, and rotation curve decompositions.

\begin{figure}
\centerline{
\psfig{figure=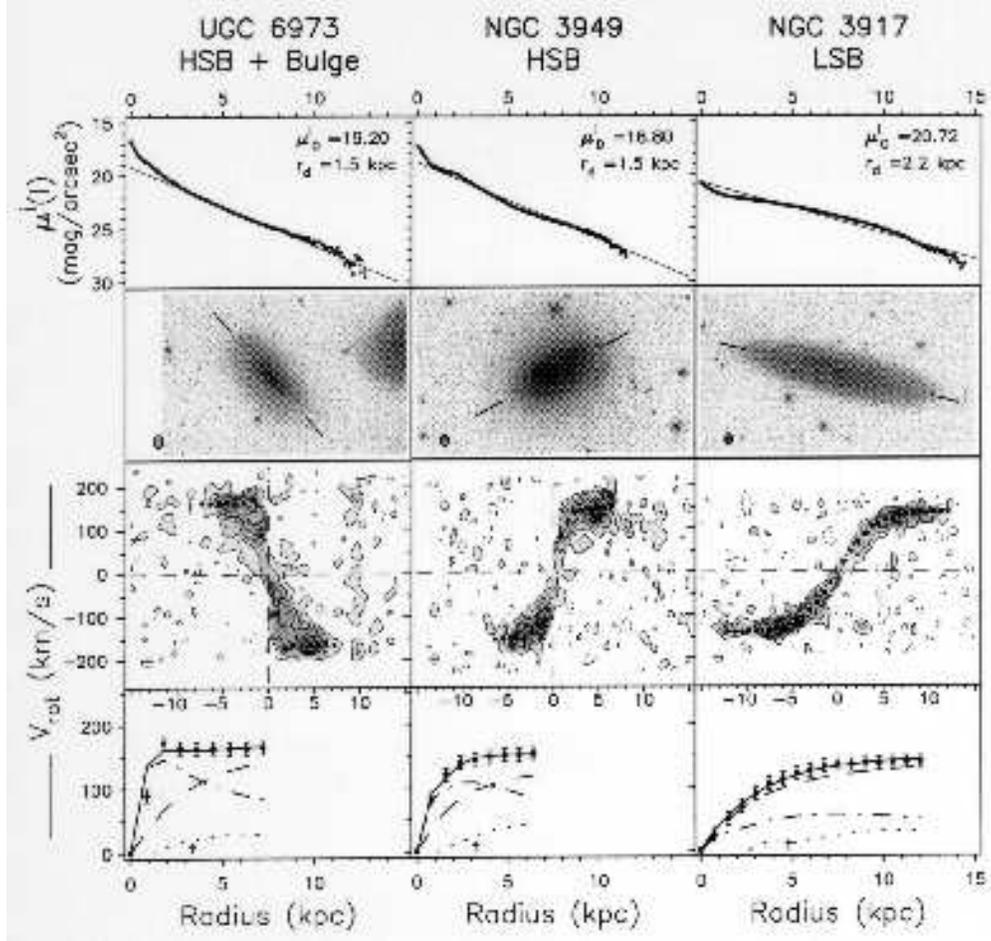,height=12.5cm}
}
\caption{
Rotation curve decompositions.  Examples of the three distinct classes
of disk galaxies are presented in each of the vertical groups.  On the
left is UGC~6973, a system with a HSB exponential disk and a central
bulge.  In the middle is NGC~3949, a system with a HSB disk but no
appreciable bulge.  On the right is NGC~3917, a system with a LSB disk
and no bulge.  In each case, the horizontal axes are position in kpc. 
In the top and bottom panels the origin with respect to the nucleus is
at the left axis, while in the middle panels the origin with respect to
the nucleus is at the center of the plots.  Surface brightnesses at $I$,
corrected for inclination, are shown in the top panels.  Images at $B$
are shown in the second row.  The major axes are indicated, as well as
the FWHM beam of the HI observations.  The velocity-position
decomposition of the HI observations is seen in the panels of the third
row.  Velocities averaged over annuli are given as dots.  The rotation
curve decompositions are provided in the bottom panels.  The dots with
error bars illustrate the observed rotation curves.  Dot--dashed curves
illustrate the amplitude of rotation expected from the observed
distributions of light and $M/L=0.4$ at $K^{\prime}$.  The dotted curves
illustrate the contribution expected from the gas component.  The dashed
curves demonstrate the contribution attributed to a dark matter halo
modelled by an isothermal sphere.  The solid lines indicate the sum of
these three components. 
}

\end{figure}

The rotation curve decompositions have three components.  One is not
important in any of these examples: the dotted curves indicate the
contributions to rotational velocities from the interstellar gas.
The dot-dash curves indicate the contributions due to the stellar
distribution with an assumed fixed $M/L$ value.  The dashed curves 
are the residual contributions associated with a dark halo.

In each case, the same value of $M/L_{K^{\prime}}=0.4
M_{\odot}/L_{\odot}$ is assumed.  This value is required for a `maximum
disk' fit (van Albada and Sancisi 1986) of the stellar components for the
two HSB examples.  Much larger $M/L_{K^{\prime}}$ values could be
accommodated for the LSB case but still a dominant dark halo would be
required.  A parallel situation is found at $B,R,I$.  It is expected
that $M/L$ values for LSBs would be lower, if anything, not higher than
for the HSBs. 
 
In the LSB case, there is the same total luminosity as in the HSB
cases, but the light is more diffuse.  If $M/L$ is fixed, then the 
mass associated with this light cannot generate such high 
velocities in the LSB galaxy.  Yet comparable rotation velocities are 
observed in the LSB and HSB examples.  For the HSB systems, reasonable 
masses associated with the observed light can explain the observed 
rotation in the inner regions.  In the LSB case, the rotation cannot 
be explained by the stellar mass and must be associated with an 
invisible component.

Tully \& Verheijen (1997) describe how a location in the 
$\mu_0, r_d$ plane of the right panel of Fig.~2 maps into a specified
ratio of the peak rotation velocity ascribed to the luminous disk,
$V_{max}^{disk}$, to the maximum observed rotation velocity, 
$W_R^i/2$, assuming $M/L=$constant.  Locii of constant values of
$V_{max}^{disk}/0.5W_R^i$ are almost horizontal lines in the right 
panel of Fig.~2.  Hence the surface brightness histograms of Fig.~1
can be transformed into the histogram of this dynamical ratio shown
in Figure~4.

\begin{figure}
\centerline{
\psfig{figure=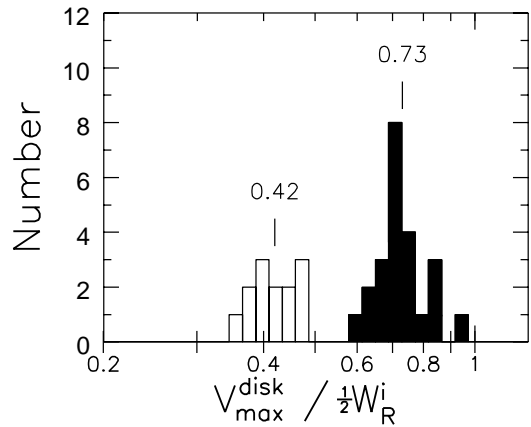,height=6cm}
}
\caption{
Histogram of the ratio $V_{max}^{disk}/0.5 W_R^i$ based on photometric 
properties.  Isolated sub-sample only (nearest important neighbor $>80$~kpc
in projection).  Values of $V_{max}^{disk}$ follow from the properties of
the exponential disk and $M/L_{K^{\prime}} = 0.4 M_{\odot}/L_{\odot}$.
The relation between $W_R^i$ and photometric parameters is given by the 
luminosity--line width correlation.  Filled histogram: HSB systems;
open histogram: LSB systems.  Mean values for each family are indicated.
}\end{figure}

\section{A Schematic Model}
The speculation is offered that this difference between HSB and LSB
systems is the rule.  In LSB galaxies, the luminous material has
settled into a disk that remains dominated by the halo potential.  In
HSB galaxies, the luminous material has become self-gravitating within
the inner 2--3 scale lengths.  Probably, the difference is a question
of angular momentum content or exchange.  High angular momentum
systems may come into rotational equilibrium in the LSB state.  Low
angular momentum systems, or parts of systems, may shrink until the
disk becomes self-gravitating.

The speculation continues that once the disk becomes self-gravitating
it must secularly evolve to find a stable configuration.  Evidently,
the intermediate surface brightness regime is disfavored.  The gap is
a factor of 10 in surface density.  The stable regime seems to abide
by the `disk--halo conspiracy' (\cite{va:sa}) which produces a flat
rotation curve.  This situation is reminiscent of the arguments of
\cite{me} for discrete states of radial stability.

\section{Acknowledgements}
Mike Pierce, Jia-Sheng Huang, and Richard Wainscoat participated in 
the collection of data.  We thank Renzo Sancisi, Erwin de Blok, and 
Stacy McGaugh for constructive comments.
This research has been supported by NATO Collaborative Research Grant
940271.
%
%

\end{document}